\documentclass{elsart}
\usepackage{graphicx,amssymb}
\journal{New Astronomy} 
\begin{document}
\begin{frontmatter}
\title{ Spectral Spatial Fluctuations of CMBR:\\ Strategy and Concept of the Experiment}
\author{Viktor Dubrovich}
\address {Special Astrophysical Observatory RAS, Pulkovskoe Chaussee 65, St.Petersburg, Russia }
\ead{dubr@MD1381.spb.edu}
\author{Anisa Bajkova}
\address {Main Astronomical Observatory RAS, Pulkovskoe Chaussee 65/1, St.Petersburg, Russia}
\ead{bajkova@gao.spb.ru}
\author{V.B.Khaikin}
\address{Special Astrophysical Observatory RAS, Pulkovskoe Chaussee 65, St.Petersburg, Russia }
\ead{vkh@brown.nord.nw.ru}

\begin{abstract}
Spectral Spatial Fluctuations (SSF) of the Cosmic  Microwave
Background Radiation (CMBR) temperature are considered as a result
of an interaction of primordial atoms and molecules with CMBR in
proto-objects moving with peculiar velocities relative to the
CMBR. Expected optimistic values of $\Delta T/T$ are $2\times
10^{-5}-2\times 10^{ - 6}$ for SSF caused by HeH$^{+}$ at z =20-30
which are possible redshifts of early reionization scenario. The
bandwidth of the lines is 0.1-2{\%} depending on the scale of
proto-objects and redshifts. For the SSF search CMBR maps in
different spectral channels are to be observed and then processed
by the Difference method. Simulation of the experiment is made for
MSRT (Tuorla Observatory, Finland) equipped with a $7\times4$ beam
cryo-microbolometer array with a chopping flat and frequency
multiplexer providing up to 7 spectral channels in each beam
(88-100 GHz). Expected $\Delta $T/T limit in the experiment is
2$\times $10 $^{-5}$ with 6$'$-7$'$ angular and 2{\%} frequency
resolution. Simulation shows that SSF may be recognized in the
angular power spectrum when S/N in single frequency CMBR maps is
as small as 1.17 or even something less for white noise. Such an
experiment gives us a possibility to set upper limit of SSF in MM
band and prepare future SSF observations.
\end{abstract}
\begin{keyword}
Cosmic Microwave Background (CMB)\sep Cosmology \sep molecules \sep Radio Telescope
\PACS 98.80.$-$k\sep 34.50.$-$s\sep 95.85.Bh
\end{keyword}
\end{frontmatter}

\section{Introduction}

Observational effects caused by primordial molecules seem to be
most promising in investigating Dark Ages epoch of the Universe.
The basic properties of the molecules are discrete narrow energy
levels and high efficiency of their interaction with CMBR. This
leads to forming SSF if proto-objects, containing these molecules,
move with peculiar velocities V$_{p}$ relative to CMBR
(Dubrovich,1977,1982, Puy,1993). We  may consider SSF as
manifestation  of the proto-objects at high redshifts 300 $> z >$
10 as scattering of CMB photons by primordial molecules must leave
imprints on the CMBR temperature distribution.

The most abundant chemical elements predicted  by the pure Big
Bang model are H, He, D, $^{3}$He, Li and their ions. They give
some molecules in the primordial matter at z=100-200 such as
H$_{2}$,H$_{2}^{+}$,HD, HD$^{+}$, HeH$^{+}$, LiH, LiH$^{+}$,
H$_{2}$D$^{+}$. There are other molecules which should be
considered if non-standard nuclear synthesis at the early times
(z=10$^9$) or star formation at z = 100-200 took place.

For the high efficiency of interaction between  molecules and CMBR
photons two values are important - the cross-section for
scattering and the concentration (or relative abundance) of this
molecule. The first parameter depends on the specific quantum
structure of the molecule - its symmetry and dipole moment. The
second one depends on the abundance of the chemical elements of
which it is composed and on the rate of the appropriate chemical
reactions. According to these constraints, we should take into
consideration only those molecules which have a large enough
dipole moment and relatively high abundance. These are HD$^{+}$,
HeH$^{+}$, LiH, H$_{2}$ D$^{+}$. We consider them as the basic
molecules interacting with CMBR at high z (Dubrovich,1977,1982,Puy
et al.,1993,Stancil et al.,1996).

\section {Theoretical predictions of SSF}

Let us remind physical conditions in Dark Ages  epoch when the
Universe was younger than 300 mil. years. The Universe was in
$(z+1)^{3/2}$ times smaller and in $(z+1)^3$ times more dense than
now, temperature of CMBR and molecular gas followed $T_{z=0}(z+1)$
law till z $\approx $150. Then, temperature of matter should be
lower than CMBR temperature due to different adiabatic indexes of
radiation and matter (Varshalovich, Hhersonskii,  Sunyaev, 1981)
and it must follow $T_{z=0}(z+1)^2$ law till first star formation.
Calculated temperatures of CMBR and molecular gas in Dark Ages
epoch are given in Table 1, we used $T_{z=0}$ = 2.726 K (Mather et
al., 1994). In epoch before secondary ionization molecules are
formed in accordance with kinetic rate of specific chemical
reactions when matter and radiation are cooling. After
reionization molecular formation follows a more complicated law
due to growth of ionization degree of hydrogen and existence of
heavy chemical elements produced by first stars (Basu,
Hernandez-Monteagudo, Sunyaev, 2004).

\begin{table}
\caption{Temperatures of CMBR and molecular gas in Dark Ages epoch z=300-10}\label{tab1}

\begin{tabular}{lllll} \hline
z & Temperature of & Temperature of & Commentary & \\
& CMBR, & molecular gas, & & \\
& K & K & & \\ \hline
1100 & 3000 & 3000 & Recombination & \\
300 & 815 & 815 & Beginning of Dark Ages & \\
150 & 415 & 415 & Middle of Dark Ages & \\
30 & 85 & 17 & Dark Ages + reionization & \\
20 & 57 & 8 & Dark Ages + reionization & \\
10 & 30 & 100-10000 & End of Dark Ages, first stars and quazars & \\ \hline
\end{tabular}
\end{table}

There are two mechanisms responsible for SSF  formation: simple
scattering (Dubrovich,1977) and luminescence (Dubrovich,1997). In
the simple mechanism opacity in narrow molecular lines and
peculiar motion of the proto-clouds results in CMBR disturbances.
This can be more visible in the rest frame of the proto-object.
The CMBR in this frame becomes non-isotropic and out of thermal
equilibrium. On the side towards which the proto-cloud moves, the
temperature of the CMBR will be higher than average and on the
opposite side it will be lower. After reflection, the photons are
distributed isotropically in this frame that leads to
non-isotropic distribution in laboratory frame. This explains the
principal role of opacity and peculiar velocity. The amplitude of
the effect depends on spectral index of the reflecting radiation
(Dubrovich, 1977, Maoli et al., 1994). This effect corresponds to
the elastic scattering between the molecules and photons, i.e. the
total number of the photons does not change.

In fact all molecules have quite a complicated  energy level
structure. This allows for the possibility of a non-elastic
process. It is the well known luminescence process which plays an
important role in the formation of radiation from a reflection
nebula. Luminescence may enhance the intensity of rotation lines
due to breaking of vibrational line quants (Dubrovich, 1997).

The magnitude of the CMBR temperature fluctuations  due to the
pure reflection of photons by a moving object depends on peculiar
velocity $V_p$ of a proto-object and optical depth $\tau$(Sunyaev
and Zel'dovich, 1970, Dubrovich, 1977, Maoli et al., 1994). The
value of the distortions produced by molecules in CMBR is:
\begin{equation}
\Delta T/T = (V_{p}/c) \cdot \tau \cdot K,
\end{equation}
where $c$ is the speed of light and $\tau $ is optical depth of
the proto-object ( $\tau << 1$, $V_{p}/c < 10^{ - 3})$, K - gain
coefficient due to  luminescence effect, K=1-1000(K=1 for simple
scattering). For the linear stage of perturbation evolution $V_p$
could be estimated as (Zel'dovich, Novikov, 1975):
\begin{equation}
V_{p}=V_{p0}/(z+1)^{1/2}
\end{equation}
where $V_{p0}$=600 km/s at large scales of CMB anisotropy (CMBA) and
factor 2-5 higher at small CMBA scales.

Luminescence effect is tempting for SSF  search with limited
sensitivity but high K may be expected  in narrow z intervals
(Dubrovich,1997) while the simple mechanism of SSF formation
operates in the wide field of z.

The optical depth $\tau$ of the proto-object  may be calculated as
(Dubrovich, Lipovka, 1995):
\begin{equation}
\tau=8 \pi^{5/2}d^2(1+z)^{3/2}n_{H}\alpha \cdot [1-exp^{-h \nu /kT}]/3hH_0,
\end{equation}
\begin{equation}
h\nu /kT=h c/[k T_0 \lambda_0(1+z)]
\end{equation}
$H_0$ - Hubble constant, h-Planck constant,  k-Bolcman constant,
$\lambda_0$ - the rest wavelength, $T_0$ - CMBR temperature,
$n_H=n_{0H}\cdot \beta $, where $n_H$ - density of $H$, $n_{0H}$ -
critical density of atomic hydrogen HI($10^{-5}$), $\beta$ - a
real density of $H$ in proto-object which may be less or much more
than unity ($\beta \ge4\cdot 10^{-2}$), $d$ - dipole moment,
$\alpha$ - abundance, $\alpha = 10^{-8}-10^{-17}$ for different
molecules and conditions in Early Universe.

Let us consider LiH, H$_{2}$D$^{ + }$, HeH$^{ + }$ HD$^{ + }$ primordial
molecules.

\underline{LiH} is a very important molecule, because it consists
of primordial Li. Its abundance is a good test for the epoch of
nuclear synthesis in the Early Universe. Its large dipole moment
and relatively low frequency of the rotational and rovibrational
transitions lead to a high value of K. But, unfortunately, some
difficulties with the chemical processes of forming this molecule
lead to a not too optimistic prediction for its abundance
$\alpha=10^{-11}-10^{-17}$. The rest wavelengths of LiH are
$\lambda_r$=676 $\mu$m(rotational transition), $\lambda_v$=7.1
$\mu$m (rovibrational transition).

\underline{HD$^{+}$} is also an important molecule due to the
presence of primordial deuterium D. The abundance of D is about 5
orders of  magnitude larger than that of Li. But HD$^{+}$ has a
dipole moment about 10 times less than LiH and a cross-section
which is 100 times smaller. Another small factor is the abundance
of H$^{+}$ at redshift z = 200, which might be about relative to
that of neutral hydrogen. Due to the relatively high frequency of
the rotational and rovibration transitions, the resulting value of
K is not very large. Yet, if high sensitivity were reached, this
molecule might be seen. The rest wavelengths of HD$^{+}$ are
$\lambda_r$=227 $\mu$m, $\lambda_v$=3.0 $\mu$m.

\underline{HeH$^{+}$} does not have any low abundance species.
There are only two small factors which lead to a low abundance: a
high rate coefficient for destruction (by electron recombination
and collisions with the neutral atoms of hydrogen) compared with
the very small rate coefficient of formation, and small abundance
of $H^+$ at high redshift. It might be the most likely molecule to
be searched for due to more optimistic predicted abundance $\alpha
= 10^{-8}-10^{-15}$. The rest wavelengths of HeH$^{+}$ are
$\lambda_r$=149 $\mu$m, $\lambda_v$=3.3 $\mu$m.

\underline{H$_{2}$ D$^{+}$} is the simplest triatomic molecule
with a high dipole moment. It contains primordial deuterium D. Due
to the presence of very low frequency transitions in its spectrum
the value of K can be very high. It is very important that the
redshift of the H$_{2}$ D$^{+}$ recombination is relatively high.
The rest wavelengths of H$_{2}$ D$^{+}$ are $\lambda_r$=1920
$\mu$m, $\lambda_v$=4.5 $\mu$m.

The expected abundances of these molecules in  the Early Universe
are discussed by many authors (Lepp and Shull,1984, Puy et al,
1993, Palla et al, 1995, Maoli et al, 1996, Stancil et al, 1996).

One of the main observational effects caused by  expansion of the
Universe is the redshift of photons, expressed as \textbf{
$\lambda $}$_{obs}$ = (z + 1) \textbf{$\lambda $}$_{em}$, where
\textbf{$\lambda $}$_{em}$ is the emitted or rest wavelength of
the photon. So, if an object contains a gas of any molecules and
has a $z$ redshift , it can only be seen at different but discrete
frequencies $\nu _{i}=\nu _{oi}$/(1 + z), where $\nu_{oi}$ is a
discrete set of the molecule's rest transition frequencies. And
vice versa, if some object is seen at given frequency $\nu _{i}$
this object can have the redshifts z$_{i}=\nu_{oi}$/$\nu_i$ -1.

The final spectrum depends on the angular size of the object
$\theta $. The angular size $\theta $ and the frequency bandwidth
$\Delta \nu/\nu $ depend on linear size L (for a spherical shape
object) and z of the object to be observed. For small z the
angular size of the object(if the linear size is constant and
equal to L) is decreasing with moving away, but it grows for z
larger than some value. Exact relations between $\theta$ and $z$
in this model contain the $\Omega_m$ and $\Omega_{\Lambda}$
parameters, which are the ratios of the matter $n_m$ and "vacuum"
densities to the critical density $n_c$ respectively
($\Omega_m+\Omega_{\Lambda}=1$). The fact that the cloud occupies
the interval of redshifts z means that if it radiates or reflects
the radiation locally in sufficiently narrow lines, all the
radiation occupies the frequency interval\textbf{ $\Delta $}$\nu
$, moreover \textbf{$\Delta $}$\nu $/$\nu $= \textbf{$\Delta
$}z/(1 + z). Scattering by molecules in several lines leads to
superposition of images of different proto-objects at different z.
Taking into account {$\Delta $}z dependence on L and $\Omega$
(Sahni, Starobinsky, 2000) we have:

\begin{equation}
\Delta \nu/ \nu\approx \theta \phi(z)[\Omega_m(1+z)^3+\Omega_{\Lambda}]^{1/2}](1+z)^{-1},
\end{equation}
\begin{equation}
\phi(z)=\int_0^z 1/\sqrt{\Omega_m(1+z)^3+\Omega_{\Lambda}}dz,
\end{equation}

\begin{equation}
\theta =(H_0 L/c)(1+z)/\phi(z),
\end{equation}
\begin{equation}
\Delta \nu/\nu\approx (H_0 L/c)\sqrt{\Omega_m(1+z)^3+\Omega_{\Lambda}}.
\end{equation}

If we simultaneously observe CMBR in two receiver  channels with
frequency difference $\Delta \nu $, the SSF with an angular size
larger than some critical $\theta_c $ are seen at both receiver
channels. If the size of the fluctuations is smaller, they are
only seen in one of the receiver channels, another one will detect
background noise only (Dubrovich,1982). In this case the
correlation function of these two observations will be within
noise to zero at small scales and different from zero at angular
scales greater than $\theta_c $. Then the difference signal of the
two channels is not zero at the scales less than
$\theta_c$($L_c$). Frequency separation $\Delta \nu/\nu$ of
receiver channels determines a critical scale of proto-objects
$L_c$ and scales $L < L_c$ remain in difference of two channels.
Frequency resolution or the channel bandwidth $\Delta \nu/\nu$
determines the optimal scale $L_{opt}$ which has the best S/N at
given z. The channel bandwidth should be a few times less than
frequency separation and $L_{opt} < L_c$.

The above consideration is correct for objects  freely expanding
over the Universe in accordance with the Hubble law. But at the
sufficiently late stages the deceleration of their expansion takes
place and contraction begins due to self-gravitation. In this case
the width of the line can be considerably smaller, and the
amplitude larger (Zel'dovich, 1978).

Fig.1. shows expected $\tau $($\alpha$, $n_H$,z),  $\Delta
T/T$($\alpha,n_H,z$), $\phi(\Omega_m, \Omega_L,z)$, $\theta(L,z)$,
$\Delta \nu /\nu (L,z)$ for HeH$^{+}$ ($\lambda_0$=1.49 10$^{-2}$
cm, d=1.66 deb ) and K=1. According to Fig.1 an optimistic level
of $\Delta T/T=2\times10^{-5}-2\times10^{-6}$ may be achieved with
$\alpha =10^{-9}- 10^{-10}$.  Expected observational frequencies
f(z) are given in Fig.1. for rotational transitions(I=0) of LiH,
H$_{2}$D$^{ + }$, HeH$^{ + }$ HD$^{ + }$. For $\phi$($\Omega_m$,
$\Omega_L,z)$ calculation we used $\Omega_m$ and $\Omega_L$ which
are discussed in several Cosmological models and recent CMBR
experiments (Ruhl et al, 2004, Miller, 2004).

\begin{figure}
\caption{$\phi(\Omega_m, \Omega_L,z)$, $\tau (\alpha, n_H,z)\theta(L,z)$, $\Delta \nu /\nu (L,z)$ $\Delta T/T$, $f(z)$ for HeH$^{ + }$}
\includegraphics[scale=1]{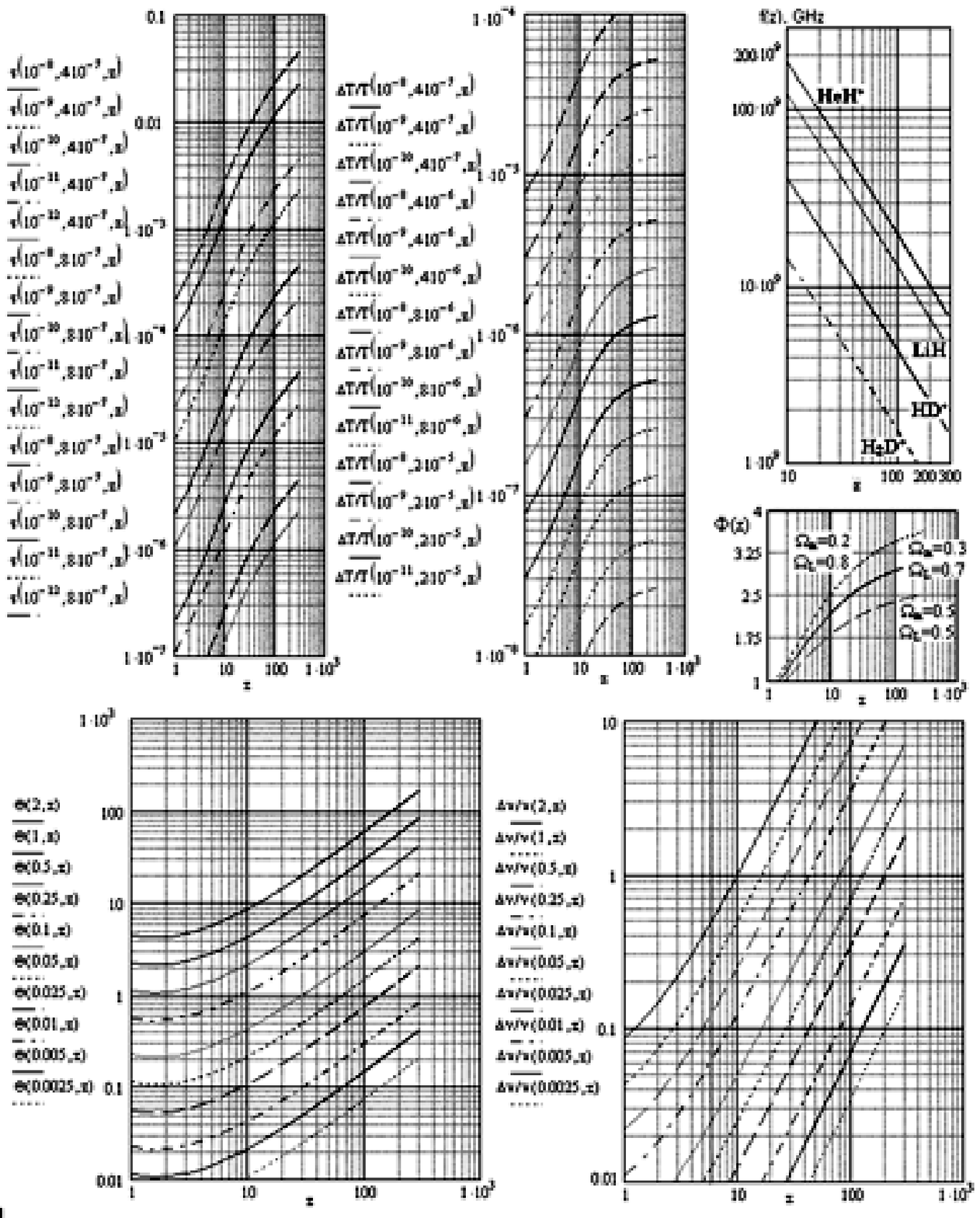} 
\end{figure}

Let us limit possible physical proto-object scales L  in the Early
Universe: 2 Mpc $>L>$0.002 Mpc. The most of Cosmological molecules
considered may exist at z$<$200 and some at z$<$300. Then possible
angular scales $\theta$ for different $\Delta \nu/\nu$, L and z
are calculated with (7),(8) and presented in Table 2. Table 2
shows that modern angular scales $\theta=5'-10'$ are still
reasonable to search for large scale SSF at z=20-30 with $\Delta
\nu/\nu$=1-2$\%$. Possible observational frequencies $f(z)$ for
HeH$^{ + }$(I=0) are given in Table 3. Most of primordial
molecules can radiate at several rotational transitions I that
leads to SSF formation at set of frequencies $f_{i}(z)$. LiH has
radiating transitions at I=0,1,2,3,4,5,6,7. The optical depth
$\tau$ in molecular lines depends on I and in the case of
thermodynamic equilibrium for optically thin layer is
(Dubrovich,1982):

\begin{equation}
\tau_{I,I+1}= \tau_0 \cdot (I+1)\cdot exp(-T_0I(I+1)/T)
\end{equation}
where optical depth $\tau_0$=$\tau$(I=0) is given by (3), $T_{0}$
-  temperature correspondent to energy of transition I=0-1, T-
temperature of radiation at given z. For LiH extremum of $\tau$
lies at I=6 ($\lambda_0 =0.95 \cdot 10^{-2}$ cm) and $\tau (I=6)/
\tau_0 $=2-4. Possible observational frequencies $f_i(z)$ for
LiH(I=6) are given in Table 4. Expected $\Delta T/T$ level for SSF
caused by LiH  with existing estimates of LiH abundance is less
than $10^{ - 5}$ that can not be achieved in this experiment but
we can set an upper limit of $\Delta T/T$ for LiH at z=30 as well.

\begin{table}
\caption{Angular and spatial scales for HeH$ ^{ + }$(I=0)}\label{tab1}

\begin{tabular}{lllllllll} \hline
$\Delta \nu/\nu,$\% & $z_{min}$ & $z_{max}$ & $L_{max}$,Mpc &
$L_{min}$,Mpc &$\theta_{max}(')$ & $\theta_{min}(')$ \\ \hline
0.1 & 5 & 200 & 0.5& 0.0025& 1.4 &0.14 & &\\
0.5 & 10 & 200& 1&0.015&4.3 &0.85 & & \\
1 & 10 & 200& 2&0.025 & 8.6 & 1.4 & & \\
2 & 15 & 200& 2&0.05 & 11.5 & 0.3 & & \\
4 & 25 & 200& 2& 0.1& 17.3 & 5.7 & & \\
5 & 30 & 200& 2& 0.15 & 20.1 & 8.5 & &\\
6 & 35 & 200& 2 & 0.17 & 22.9 & 9.6 & & \\
7 & 40 & 200& 2 & 0.2 & 25.8 & 11.3 & & \\
10 & 50 & 200& 2 & 0.25 & 31.4 & 14.2 & & \\ \hline
\end{tabular}
\end{table}

\begin{table}
\caption{ Observational frequencies f(z)in GHz for HeH$ ^{ + }$(I=0) }\label{tab1}
\begin{tabular}{llllllllllllllllllllllllllllll} \hline
\ z&5&10&15 &19&20&22&52&60&65& \\ \hline
f &335.57&183.30&125.84&100.67&95.88&87.54&37.99&33.00&30.51& \\ \hline
\end{tabular}
\end{table}
\begin{table}

\caption{Possible observational frequencies f(z) in GHz for LiH(I=6) }\label{tab1}
\begin{tabular}{llllllllllllll} \hline
\ z&10&15&20&30&32&34&35&80&101& \\ \hline
f&287.08&197.37&150.38&101.87&98.68&90.23&87.71&38.9&30.96&\\ \hline
\end{tabular}
\end{table}

\begin{table}
\caption{Spatial scales L of peaks in CMBR power spectrum for different z}
\begin{tabular}{llllllll} \hline
$z$ & $L_1$, Mpc & $L_2$,Mpc & $L_3$,Mpc & $L_4$,Mpc \\ \hline
1000& 0.22 & 0.11 & 0.073& 0.043& \\
100& 1.99 & 0.99 &0.66 & 0.39 & \\
50& 3.72 & 1.86& 1.24 & 0.74 & \\
30 & 5.79 & 2.91 & 1.93 & 1.16 & \\
20& 8.08 & 4.04 & 2.69 & 1.62 & \\ \hline
\end{tabular}
\end{table}

If we realize a channel bandwidth of $\Delta \nu/\nu$=2{\%}  in 3
mm band preferable or optimal linear scales for SSF search are
L=1.5 Mpc for HeH$ ^{ + }$(I=0) at z=20  and L=0.75 Mpc for
LiH(I=6) at z=30, correspondent angular sizes according to (7) are
10.7$'$ and 7.5$'$.

Recent WMAP polarization measurements have shown strong
probability of an early reionization scenario with
$z_{r}=20(+10/-9)$ (Kogut et al, 2003, Page et al, 2006) due to
Pop III stars (Wyithe, Loeb,2003) in Galaxy halos. SSF detection
at z=20-30 may provide us with important information about
ionization history of the Universe.

The question is which maximum proto-object scales  are reasonable,
could exist at z=20-30 or are connected with some known scales in
the Universe. Let us compare them with spatial scales of primary
CMBA. The angular scale of the first peak in primary CMBR power
spectrum $\theta_1 \approx$ 50$'$, the following CMBA peaks are
$\approx$ 1/2$\theta_1$,1/3$\theta_1$, and 1/5$\theta_1$ (see
Fig.2 down). Then possible spatial scales L of 1-4 peaks in CMBR
power spectrum for different $z$ are given in the Table 5 with (7)
and (6).

The above consideration shows that at $\Delta $T/T=2$\cdot $10
$^{-5}$  level  and with $\Delta \nu/\nu$=2{\%} we can search for
SSF which are the result of large scale proto-object motion with
peculiar velocity relative to CMBR at z=20-30. These possible
proto-objects are still less than spatial scale of the fourth peak
in CMBR power spectrum (Table 5) which might be related to a
proto-galaxy scale. With less $\Delta \nu/\nu$, higher sensitivity
and multipole resolution we can approach to smaller spatial scales
of proto-objects up to proto-star scales.

SSF firstly proposed as a new phenomena in (Dubrovich,  1977) is
one of most promising secondary CMBA. SSF is  non-isotropic
spectral distortions of CMBR which are to be present at z=300-10.
Homogeneous and isotropic spectral distortions of CMBR may be
present at z=6000-1000 due to primordial Helium and Hydrogen
recombination (Dubrovich, 1975, Dubrovich, Stolyarov, 1995,
Dubrovich, Shahvorostova, 2004, Dubrovich, Grachev, 2004).
Non-isotropic distortions of CMBR similar to SSF may be present in
this epoch at lower $\Delta T/T$ level as well(Dubrovich, 1977/2).
Calculation of this effect in more details at z=1000-1100 is done
in (Rubino-Martin, Mernandez-Monteagudo and Sunyaev, 2005).

\section{Simulation of SSF}

Let us remind that CMBR temperature fluctuations are  expanded in
spherical harmonics (Peebles, 1993,Peacock, 1999):
\begin{equation}
\Delta T(\theta,\phi)/T_0=\sum_{l=0}^{\infty}\sum_{m=-1}^l a_{lm} Y_{lm}(\theta,\phi),
\end{equation}

The angular power spectrum $C_l$ of $\Delta T/T$ can be  related
to the autocorrelation function $C(\theta)$ as:
\begin{equation}
C(\theta)=1/4\pi \sum_l(2l+1)C_l P_l(cos \theta),
\end{equation}

where $P_l$ are Legendre polynomials, l-multipole moment and
$C_l=<\vert a_{lm}\vert ^2>$. Theoretical predictions of $\Delta
T$ are usually quoted as

\begin{equation}
(\Delta T_l)^2=l(l+1)C_l/2\pi,
\end{equation}
\begin{equation}
\Delta T_l=\sqrt{l (l+1) C_l/2\pi},
\end{equation}

As we consider small-angle patches of the sky  CMBR fluctuations
$\Delta T = T- <T>$ may be generated by evaluating the simple
Fourier series (Bond et al, 1987):

\begin{equation}
\Delta T(\theta_x, \theta_y)/T
=\sum_{n_u=0}^{N_u-1}\sum_{n_v=0}^{N_v-1} D(n_u,
n_v)\exp[i\frac{2\pi}{L}(n_u\theta_x + n_v\theta_y)]
\end{equation}
where $L$ denotes the size of the simulated region in  radians;
$(\theta_x, \theta_y)$ are Cartesian coordinates on the sky
(spatial domain); $(n_u,n_v)$ are coordinates of Fourier
components $D$ in spatial frequency domain, $<\vert
D(n_u,n_v)\vert^2>=C_l$, $l=2\pi/L\cdot \sqrt{n_u^2+n_v^2}$.

For the search of SSF CMBR maps obtained in several spectral
channels are to be studied and then processed by the special
frequency-differential or the difference method (Dubrovich,
Bajkova,2004). The Difference method can be considered as an
alternative method to the correlation function analysis. In this
case we are interested in the first derivative of spatial
distortions with respect to the frequency. The method used is
based on the analysis of a difference of two CMBR temperature maps
observed at different frequencies and reduced to one beam shape.
Mainly, such a difference map contains information only on the
secondary fluctuations because the primary CMBR fluctuations
present in both maps will be eliminated due to their black-body
spectrum nature.

The procedure of SSF simulation is the following.  Firstly CMB map
is generated with a standard CMB power spectrum. Then  amplitudes
of FFT spectrum of CMB map are kept while phases are modified as
random values uniformly distributed in the interval 0-2$\pi$. A
backward FFT gives us an approximation of CMB map which contains
primary and secondary fluctuations in conditions of unknown
theoretical power spectrum of SSF.  The actual theoretical  power
spectrum of SSF is to be  a  subject of nearest energetic efforts,
it is strongly needed for following SSF investigation.

Let the difference of the CMBR observation frequencies be equal to
$\Delta\nu_1$. Let the limiting angular fluctuation size
$\theta_1$ corresponds to this frequency difference in accordance
with equation (5). Obviously, the fluctuations of size larger than
$\theta_1$ will be seen at both maps. After subtraction of one map
from the other the fluctuations of size larger than $\theta_1$
will be mutually suppressed and the remainder map will contain
only the fluctuations smaller than $\theta_1$. Evidently, this
fact can be seen also from angular power spectrum of this map
which is close to zero at frequencies with multipoles $l<l_1$ and
different from zero at $l>l_1$, where $l_1\approx 1/\theta_1$ is
the break of the spectrum. Now let the difference of two
frequencies be $\Delta \nu_2$, where $\Delta \nu_2 > \Delta
\nu_1$. The angular size $\theta_2$ corresponds to this frequency
interval. Obviously, $\theta_2 > \theta_1$ in accordance with
equation (5). In this case we will see at the difference map the
fluctuations with size smaller than $\theta_2$ and the
corresponding angular power spectrum will have the break at
$l_2\approx 1/\theta_2 <l_1$.

Our CMBA model also includes convolution of expected  CMBA maps
with a simulated radio telescope beam and adding a pixel noise.
Differential maps are preliminary processed to decrease the noise
in given power spectrum region. Let convolved CMBR maps have
spatial scales l$_{min }$ $\div $ (l$_{max}$-m$\Delta l_{m})$,
then the spatial scale of differential maps is (l$_{max}$-n$\Delta
$l$_{m}$)$\div $l$_{max}$, where m=0,1,2..., $\Delta $l$_{m}$/l
corresponds to $\Delta \lambda _{m}$/$\lambda ($2{\%} in our
case). Let us limit l$_{max }$ in the simulation by l=1056 which
is near to the lowest reasonable limit of l$_{max}$ in the ground
based MM-wave SSF experiment. We will use in simulation a standard
$\Lambda $CDM CMBR model (Spergel et al., 2003, Spergel et al,
2006) with parameters: $\Omega_b\times h^2$=0.2($H_0$/100 km/sMpc,
$H_0=65$ km/sMpc), $\Omega_\lambda$=0.65, $\Omega_M=0.3$, n=1,
h=0.65. Initial single frequency CMBR maps (Fig.2 up)
15$^{o}$x15$^{o}$ (128x128 pixels) have l=3-1056(1), l=3-1032 (2),
l=3-1008 (3), l=3-984(4), l=3-960(5), l=3-936(6), l=3-912(7), all
$\sigma $(quadratic sum)=25902. Difference maps (Fig.2 down) have
l=1032-1056 (1-2), $\sigma $=14.4, l=1008-1056(1-3), $\sigma
$=21.3, l=984-1056(1-4), $\sigma $=32.4, l=960-1056(l-5), $\sigma
$=42.9, l=936-1056 (1-6), $\sigma $=60.8, l=912-1056, $\sigma
$=87.6 with the angular power spectrum corresponding to $\Lambda
$CDM model for initial (solid line) and differential (dashed line)
CMBR maps (Fig.2 middle).

\begin{figure}
\caption{ Initial single frequency CMBR maps 15$^{o}$x15$^{o}$
(128x128 pixels) (up), difference maps (middle), power spectrum of
$\Delta $T corresponding to $\Lambda $CDM model for initial (solid
line) and differential (dashed line) CMBR maps (down)}
\includegraphics[scale=1.5]{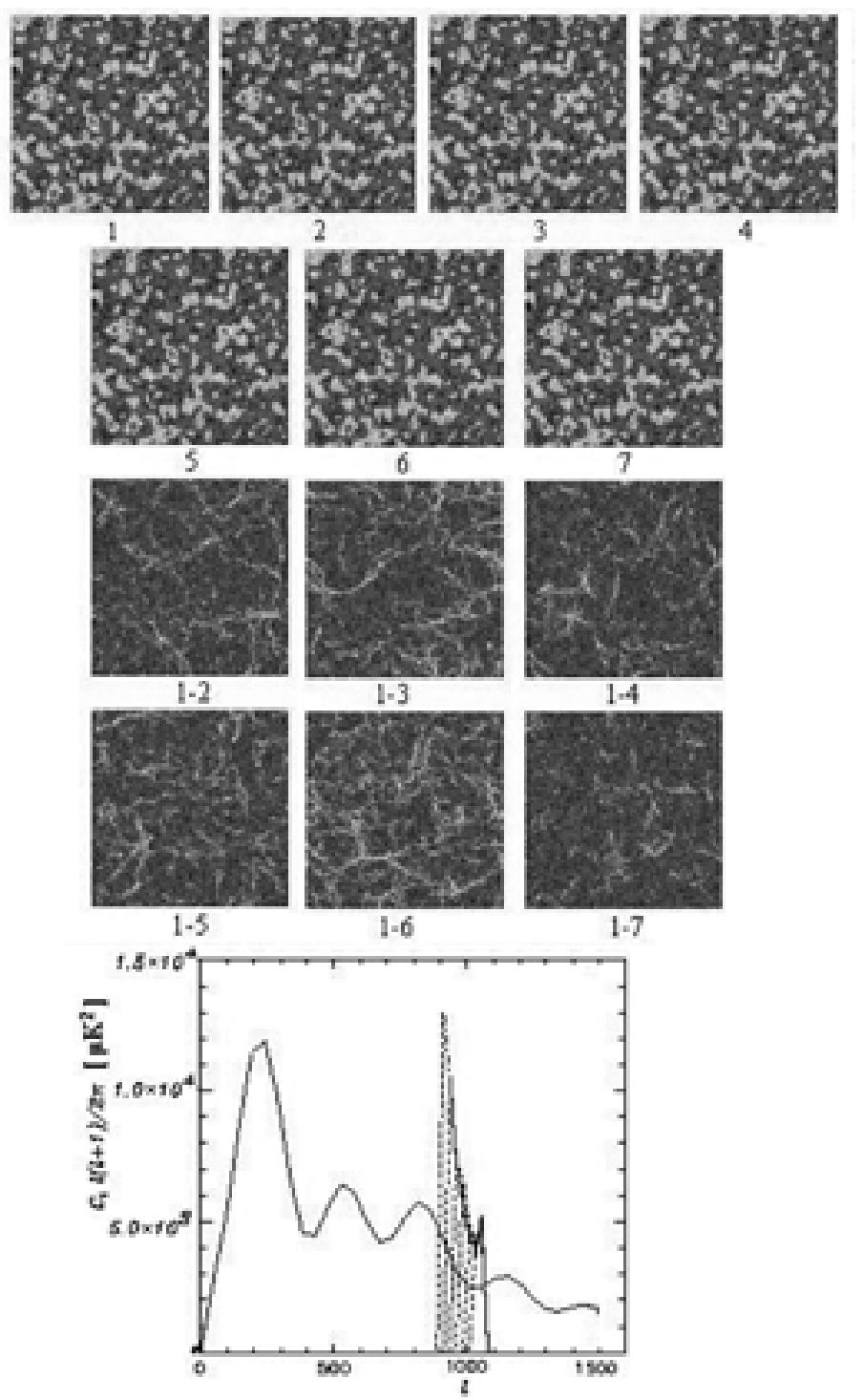} 
\end{figure}

It was shown in (Dubrovich, Bajkova,2004) that the Difference
method may be used when S/N ratio for each map is 1.44. In case of
higher level of noise a more complicated preliminary signal
processing directed to decreasing the noise in the given CMBR
spectrum region would be required. Fig.3 demonstrates predicted
difference SSF map without an instrumental noise(up) and with
white instrumental noise (down) added in single frequency maps
with S/N=1.17. Power equalization method of Wiener filtration
(Gorski, 1997) was applied in this case as signal preliminary
processing to decrease the noise in given power spectrum region.
In spite of a correlation coefficient of two difference maps of
Fig.3 is less than 0.15 the power spectra of the difference signal
after Wiener filtration are similar (Fig.3 right) that is the most
important to recognize  SSF in  a real SSF search experiment. But
this is true for white instrumental noise and we  concentrate our
efforts in section 4 to avoid or minimize abnormal instrumental
pixel noise in the SSF search  experiment.

\begin{figure}
\caption {Predicted difference SSF map without noise(left up)  and
with  white noise (left down), S/N=1.17,  the correspondent power
spectrum without noise - solid line and with white instrumental
noise - dashed line (right)} \vskip 9.5 cm
\includegraphics[scale=1]{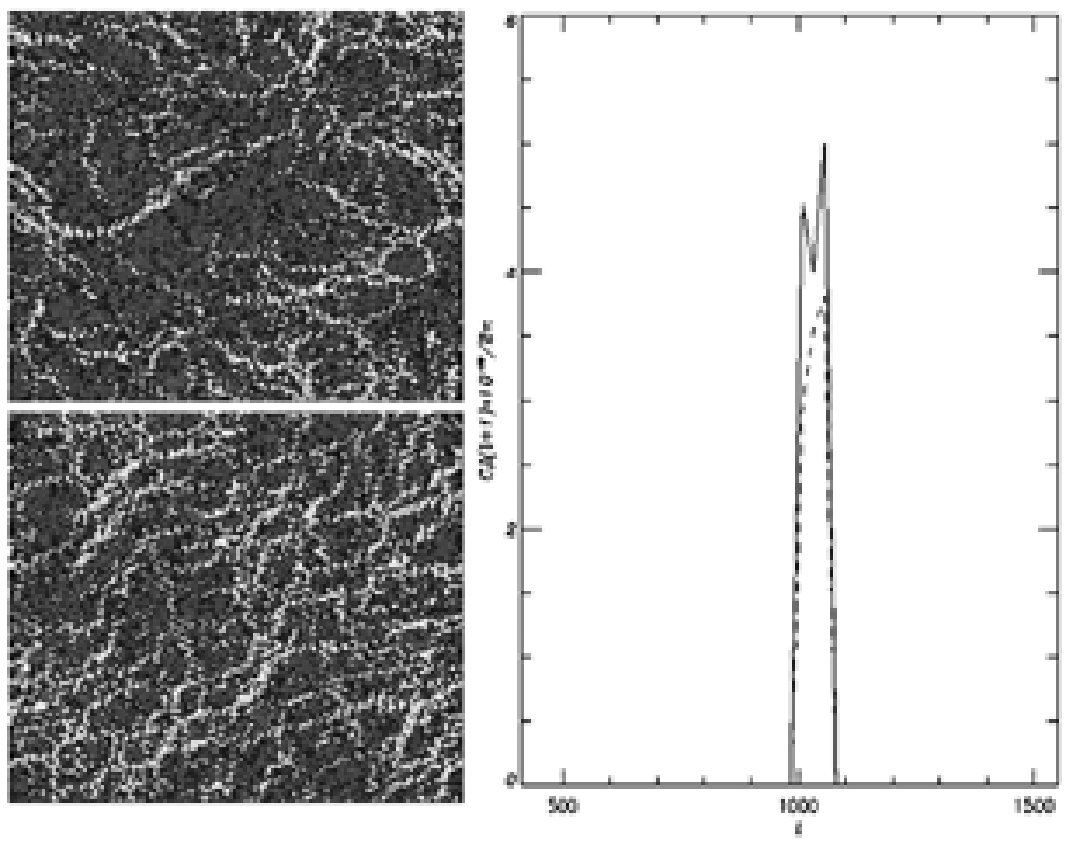} 
\end{figure}

It was assumed in above simulation that the observed CMBR maps
consist only of the primary and secondary CMBR fluctuations with
some additive white instrumental noise. However,  in practice the
CMBR detected maps in addition contain  a number of astrophysical
foregrounds such as extragalactic unresolved sources, free-free
and  synchrotron radiation, SZ effect, Galactic dust and so on.
Astrophysical foregrounds are characterized by considerably low
value of spectral index as compared with the analyzed secondary
CMBR fluctuations and are successfully eliminated in difference
maps.

\section{A concept and strategy of SSF Experiment}

Up to now  practically all CMBR ground based and  space
experiments study primary CMBA which has black body spectrum.
Primary CMBA was 30 years under investigation at RATAN-600 radio
telescope since 1977 (Parijskij, Korolkov, 1986, Parijskij et al,
2005).

First attempts of SSF observations were done in 1992 at  IRAM MM
radio telescope with the upper $\Delta $T/T limit 2$\cdot $10$^{-
3}$ at 1.3 mm (De Bernardis, Dubrovich et al, 1993) and in 2001 at
RATAN-600 (the near pole survey) with the upper $\Delta $T/T limit
10$^{ - 3}$ at 6 cm (Gosachinskii, Dubrovich et al.,2001). Surplus
frequency resolution of the existing spectrum-analyzers (30 MHz
and 20 KHz) increased $\Delta $T/T limit in both cases. The Far
Infrared Absolute Spectrometer(FIRAS) instrument has shown that
CMB has ideal black body spectrum from 60 GHz to 600 GHz and CMB
temperature is 2.725+-0.001 K(Fixsen, Mather,2002). Then Fixsen et
all reported about an effort to measure distortions of CMB black
body spectrum due to of various processes in Early Universe. This
attempt was done in ARCADE balloon-born experiment where CMB
temperatures were measured at 10 GHz and 30 GHz with accuracy
+-0.010 K and +-0.032 K (Fixsen et all, 2004).  In this paper we
suggest a concept and strategy of the SSF experiment with  $\Delta
$T/T limit  $2 \cdot 10^{-5}$. High absolute measurement accuracy
is not necessary in this experiment as we need to measure
differential temperatures in nearby receiver channels without
measuring absolute CMB temperatures.

Simulation of possible SSF experiment is done for MSRT (Khaikin et
al, 2002,2003), Tuorla Observatory, Finland (Fig.4) equipped with
a 7x4 beam cryo-microblometer array at 3 mm with RF multiplexer
providing up to 7 spectral channels in each beam. The scheme of
the experiment is given in Fig.5.  Firstly idea of such an
experiment is suggested  in (Dubrovich, Bajkova, Khaikin, 2004,
Khaikin, Dubrovich, 2004, Khaikin, Luukanen, Dubrovich, 2005). In
this paper we develop   the  concept and present simulation
results of a possible experiment.

\begin{figure}
\caption{Multibeam Solar Radio Telescope (MSRT) of  Tuorla
Observatory, Finland }
\includegraphics[clip,scale=1]{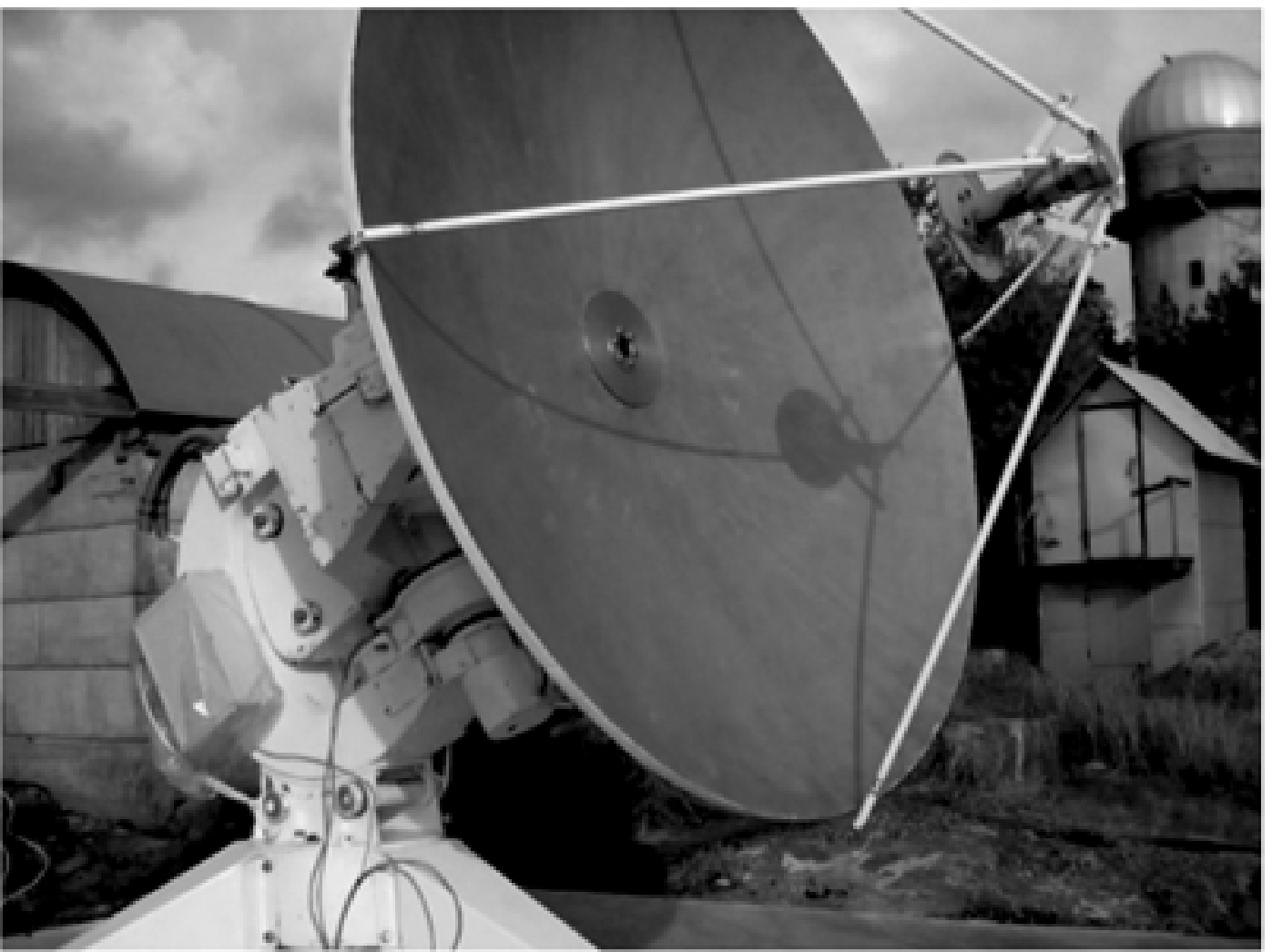} 
\end{figure}

\underline{RF filters}

The challenge is to split RF band in front of bolometers.  If
superconducting Nb microstrip resonant $\lambda $/4 stub filters
are used in RF multiplexer (RFM) the in-band transmission can
approach unity due to negligible loss of the superconducter. To
simplify a multiplexer design the number of filters and spectral
channels may be reduced to 5 or even 3. A wideband microstrip
superconducting frequency multiplexer is applied in front of TES
bolometers in SAMBA(Superconducting Antenna-coupled Multifrequency
Bolometric Array), where low loss and excellent out-of-band
rejection is demonstrated at 3 frequencies of MM band (Goldin et
al, 2003). Antenna-coupled TES bolometer with an integrated
microstrip BP filter at 217 GHz is applied for a bolometer array
in a ground based CMB polarization experiment (Myers et al, 2005).

Another way to split radio frequency band in a bolometric
receiver is to build a quasi-optical filter based on metal mesh
filters(MMF) or frequency-selective surfaces (FSS). A review of
quasi-optical MMFs is given in (Ade et al, 2006). A few possible
designs of cryogenic MMFs  with bandwidth up to 6$\%$ at 1-3 mm
are given in (Page,1994). 3 cryogenic filter wheels with several
bandpass filters,  are under development now in Columbia
University for Millimeter Bolometric Array Camera (MBAC) to
provide  5-10{\%} frequency resolution in one octave MM band from
150 GHz to 300 GHz at Atacama Cosmology Telescope(ACT).

FSS may be also used to separate in space  radio telescope  foci
in which several feeds or arrays at different frequencies may be
placed (Kasyanov, Khaikin, 2004). But it is difficult to achieve
frequency resolution with MMF or FSS better than 5$\%$ and
superconducting $N_b$ microstrip resonant filters are more
effective in our case.

\underline{Array feeds}

We consider several variants of an array feed. As a  narrow flare
corrugated horn is difficult to build in 3 mm band a trimode
conical horn may be used to provide a small flare angle with an
acceptable sidelobe level (less than --20 dB in E/H plane) and low
frequency cut-off near  87 GHz. A smooth-walled spline-profile
horn(Granet et al, 2004) is an alternative to both corrugated and
trimode horns and it allows us to achieve an identical  radiation
pattern in E/H plane in the band, low sidelobe level(less than -25
dB), more tight array packaging and lower manufacture cost. We are
in final stage now to build and test such a horn for array
applications al lower frequencies (30 GHz-38 GHz) in cooperation
with C.Granet(CSIRO)(Khaikin et al,2007).

\underline{Multipole resolution}

A 2 m precise carbon fiber MSRT dish (Khaikin et al,2002)  allows
us to work in 100 GHz and 150 GHz bands but the second one is much
more difficult due to higher atmospheric attenuation and antenna
scattering background. Angular resolution of MSRT is 7$'$-6$'$ at
88 GHz-100 GHz that is enough for a large scale SSF experiment and
corresponds to l$_{max}$=$\pi$/$\theta$=1542-1800 (for full sky
coverage) the same as at Plank (7$'$, l=1542) and twice better
than at WMAP(l$_{max}$=750). Nevertheless for a ground based or
balloon experiment $\Delta l<l<l_{max}-\Delta l$, where $\Delta
l=2\pi /M$ (size of the symmetrical map) and actual spatial
resolution of MSRT is not higher than l=1200-1500 for partial sky
coverage. In Boomerang balloon experiment actual resolution was
less than l=1025 at 150 GHz(de Benardis et al.,2002). 1000$\ge l
\le 1500$ seems to be optimal for SSF search at current stage. At
higher l with the low frequency resolution S/N ratio falls.
Besides, in ground based experiments with higher angular
resolution such as Cosmic Microwave Imager (CBI) (Readhead et
al,2004) with l $_{max}$= 3500 point sources were a serious
problem at 26-36 GHz as the power spectrum of primary CMBA falls
rapidly with l while the point source contribution grows as
l$^{2}$ in the band. Point source contamination is still dangerous
at 3 mm at higher l if secondary CMBA has a falling spectrum. Some
secondary CMBA may have inverted $C_l$ spectra when the magnitude
of $C_l$ grows at higher l (Doroshkevich, Dubrovich, 2001).
However  SSF search at l$>$1500 needs both higher sensitivity and
higher frequency resolution which contradicts each other.

\underline{Chopping mode}

A multifeed or receiver array is needed in a SSF  experiment to
get a CMBR map in an unmovable radio telescope mode (preferable)
or to accelerate mapping in a scanning mode (less preferable
because of variable instrumental effects). We plan to use a
chopping mode to reduce an abnormal instrumental noise and
atmospheric effects. A chopper with beam deviation  $\theta_{ch}$
will distort power spectrum of SSF due to filtration of scales
$\theta > \theta_{ch}$ and changing the amplitudes at scales
$\theta < \theta_{ch}$ but this is not so dangerous in the task of
SSF search. The main role of the chopper is to suppress
uncorrelated residual 1/f noise of different bolometers which can
not be eliminated in difference maps.

\underline{ Calibration}

Identical and stable calibration of all receiver  channels is very
important in this experiment. It may be produced quasi-optically
by a lens, forming a lateral plane wave for a whole array.  A high
stable noise generator with a direct coupler  or a matched black
body load is placed in the lens focus. A calibration source must
be put in a reverse two level thermostat realizing 0.1 K accuracy
at the first and not less than 0.01 K accuracy at the second
level.

\begin{figure}
\caption{ Suggested scheme of the experiment (up), multichannel
bolometer receiver (MBR) scheme with RF multiplexer (down)}
\includegraphics[clip,scale=1]{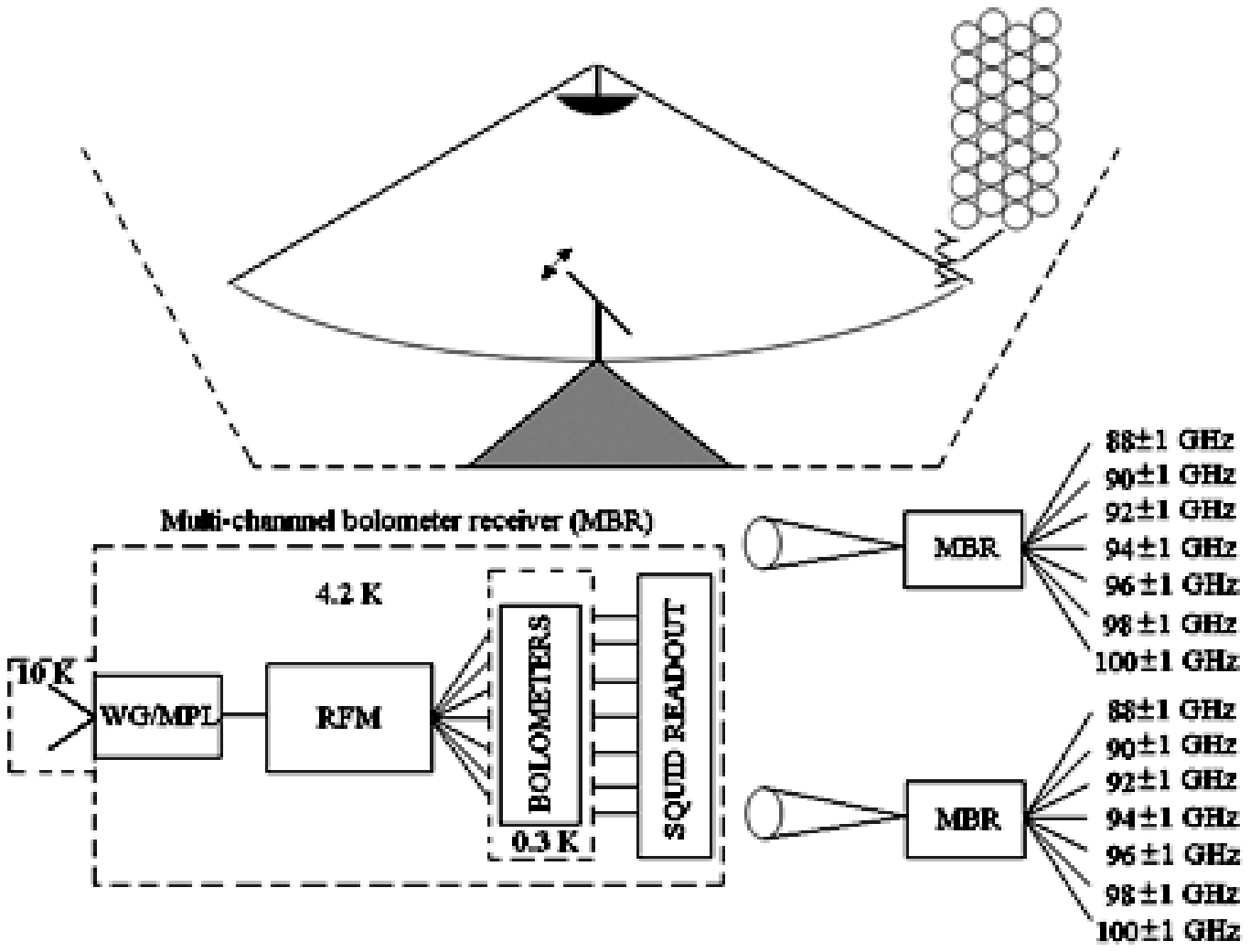} 
\end{figure}

\underline{CMBR map}

According to our concept MSRT central beam is  fixed near the
North Celestial Pole (NCP) at $\delta $=86$^{o}$ to increase
cosmic source passing time via the beam up to 348s -402s and to
observe the 1$^{o}$ x 8$^{o}$  strip preferably in low culmination
from $\alpha $=8$^{h}$00$^{m}$ to 16$^{h}$00$^{m}$ at high
Galactic latitudes. Actual spatial resolution of a combined CMBR
map in the scheme of Fig.5 is 6$'$-7$'$(HPBW). The long-length
focus Cassegrain scheme with m=(e+1)/(e-1)=10-11 (eccentricity
e$>$1) gives us negligible aberrations of off-axis beams but
requires more complicated array feeds. A short-length focus
Cassegrain radio telescope scheme gives us unacceptable level of
off-axis aberrations(Khaikin, Luukanen, 2003).

\underline{Antenna characteristics}

We simulated some antenna characteristics for  the scheme of the
experiment given  at Fig.5. Antenna parameters are D=2 m, d=0.3 m,
F/D=0.435, interfocal distance f=150cm, feed flare angle is 12
deg. Simulated characteristics are given in Table 6.  Multibeam
Focal Plane Array (FPA) configuration presented in Fig.5 allows us
to avoid undersampling in a survey mode. For simulation of the
multibeam radiation pattern, calculation of some other antenna
characteristics (illumination, aperture, spillover, sidelobe
efficiencies, beam overlapping level  etc.) FOcal Plane Array
Simulation (FOPAS) program package has been used (Khaikin, 2005).
FOPAS takes into account  geometry of a single or dual reflector
radio telescope, expected aperture distribution or the beam
pattern of an ideal broadband feed, the feed removal from the
focus and off-axis aberrations. FOPAS allows  us   to  optimize
FPA  geometry  with the beam  spacing and the overlapping level
which are optimal  for the given astrophysical task. The task of
multibeam FPA optimization is also considered in (Majorova,
Khaikin, 2005). Multibeam radiation pattern simulated with FOPAS
is given in Fig.6.

\begin{table}
\caption{ Simulated antenna characteristics}\label{tab1}

\begin{tabular}
{|p{50pt}|p{50pt}|p{60pt}|p{60pt}|p{60pt}|}
\hline
\textsf{Number } \par \textsf{of beams}&
\textsf{Feed taper,} \par \textsf{dB} \par \textsf{}&
\textsf{HPBW at} \par \textsf{88-100 GHz,} \par \textsf{arcmin} \par \textsf{}&
\textsf{FSL,} \par \textsf{Db}&
\textsf{Beam } \par \textsf{overlapping} \par \textsf{level,} \par \textsf{dB} \par \textsf{} \\
\hline
\textsf{28}&
\textsf{-10/-13}&
\textsf{7-6}&
\textsf{21/24}&
\textsf{-0.45} \\
\hline \textsf{Spillover} \par \textsf{efficiency,} \par
\textsf{dB} \quad & \textsf{Blockage} \par \textsf{efficiency}
\par \textsf{dB}& \textsf{Illumina-} \par \textsf{tion
efficiency}& \textsf{Ruze efficiency (94 GHz)}&
\textsf{Antenna efficiency (94 GHz)} \\
\hline
\textsf{-0.45}&
\textsf{-0.23}&
\textsf{0.91}&
\textsf{0.75}&
\textsf{0.51} \\
\hline
\end{tabular}
\end{table}

\underline{TES bolometers}

We consider Transition Edge Sensors (TES) as cryo-bolometers for
this experiment (Khaikin, Luukanen, Dubrovich, 2005). The lateral
dimension of TES is less than 10 microns (microbolometer) and
standard lithographic techniques may be used to build arrays of
100-1000 sensitive elements.

\begin{figure}
\caption{Simulated multibeam radiation pattern of MSRT at 100 GHz,
the feed taper is -13 dB}
\includegraphics[clip,scale=1]{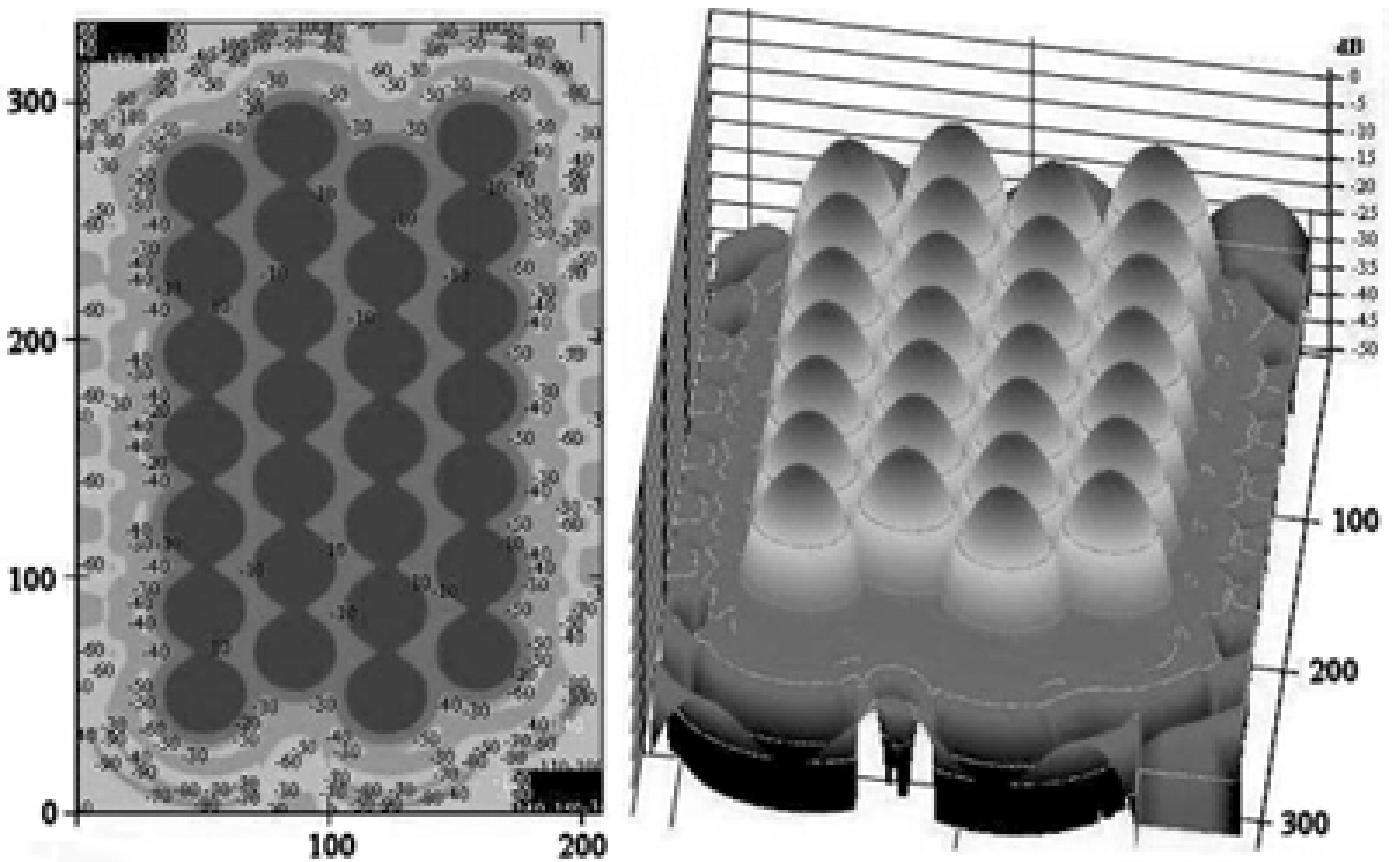} 
\end{figure}

Photon and phonon Noise Equivalent Power (NEP) and Noise
Equivalent Temperature Difference (NETD) of TES for the ground
based SSF experiment are considered in (Khaikin, Luukanen,
Dubrovich, 2005). An expected phonon NEP of a cryo-microbolometer
with the physical temperature of 0.3 K is $<$1$\cdot$10$^{-17}$
W/Hz$^{1/2}$. We do need the physical temperature near  0.3 K
(Fig.5) as NEP of TES dramatically grows at higher physical
temperatures (Luukanen, Pekola,2003). Photon NEP from radio
telescope antenna and atmosphere is not less than
2$\cdot$10$^{-17}$ in good atmospheric conditions. This will give
us $\sigma $=1.0-2.0 mK s$^{1/2}$ sensitivity per beam per channel
(2 GHz bandwidth) in the best atmospheric conditions with
T$_{atm}$$<$20 K (the lowest atmospheric opacity). In 8x8 element
planar bolometer array at 90 GHz for GBT operating temperature is
450 mK, which limits NEP of TES bolometers to 2$\times$10$^{-17}$
W/Hz$^{1 / 2}$ (Benford et.al,2004). The physical temperature
$T_p$ of an antenna coupled with a superconducting bolometer must
be chosen to provide a photon NEP$<$10$^{-17}$ W/Hz$^{1 / 2}$. For
rather high radiative antenna efficiency $\eta $=0.9 and
T$_{p}$=300 K an additional photon noise from an antenna is 10$^{
- 16}$ Wt/Hz$^{1/2}$  and cooling of antenna (horn)up to at least
10 K is necessary to provide its own photon  NEP at less than
10$^{- 17}$ Wt/Hz $^{1/2}$ level (Khaikin, Luukanen,
Dubrovich,2004)(Fig.5). For deep cryogenic cooling Sorption $He^3$
coolers at 0.3 K of Applied Physics Institute of RAS in Nizhnij
Novgorodmay be used(Vdovin, 2005).

Some ground based CMBR projects (Lawrence et al,2004) chose  an
alternative way and  are being built cryogenic MMIC MM-wave
cameras with expected sensitivity of 500 $\mu K s^{1/2}$ in
continuum in 3 mm band. We have considered such alternative as
well. InP MMIC LNA at 15 K with a broadband subharmonic mixer and
IF microstrip 10 channel  multiplexer may give us 4-5 mK$s^{1/2}$
per beam per channel(1{\%} bandwidth) at 85 GHz -105 GHz (Khaikin
et al, 2004).

\underline{Sensitivity in the experiment}

The actual reachable integration time of a bolometer depends on
the Knee frequency f$_{kn}$ in its noise power spectrum. In a
bolometer case f$_{kn}$ may be reduced from tens of Hz to tens of
mHz by biasing the bolometer with an AC current and synchronous
demodulation in readout electronics (Crill et al, 2002).
f$_{kn}$=10 mHz was reached by AC biasing of bolometers in a 144
pixel cryo-bolometer array BOLOCAM at 140-270 GHz (Mauskopf et
al., 2000). With AC biasing of the TES bolometers we expect
f$_{kn}$=20 mHz. In the effective chopping mode with chopper
frequency f$_{ch}$=5 Hz  f$_{kn}$=5 mHz may be reached then 3-4
minutes integration time becomes available. Using a 100 ms
sampling rate for f$_{ch}$=5 Hz and 170 s synchronous integration
time we can reach $\sigma $ of thermal noise not worse than
0.1-0.2 mK per beam per pixel. Four 1$^{o}$x8$^{o}$  maps (128x7
pixels)  are to be recorded during about 8 hours in clear cold
winter nights (after Sunset and before Sunrise) and then combined
into one map (64x14 pixels) with minimum angular resolution
$6'\times 6'$ and a noise factor of 1/2$^{1/2}$ in comparison with
initial maps. 20-25 identical maps must be summed for 20-25 clear
cold nights with low atmospheric opacity in order to reach
30$\mu$K $\sigma $ of thermal noise in the averaged single
frequency maps. We expect contribution of non thermal instrumental
noise, atmosphere and foregrounds in differential maps up to 30
$\mu$ K as well. So we may expect total $\sigma $ =50-55 $\mu$K in
differential maps.

\underline{Atmospheric limitations}

Big efforts have to be made to reduce  the contribution of
instrumental and foreground effects at $\Delta $T/T=10$^{-5}$
level that corresponds to 10$^{-7}$ of ambient temperature. Some
instrumental effects such as 1/f noise, ground spillover,
radiometric offset may be comparable with an atmospheric noise but
their contribution may be significantly reduced in the ground
based experiment. Atmospheric fluctuations are to be effectively
suppressed  with a chopper while high atmospheric attenuation is
the most dangerous for a ground based MM-wave SSF experiment with
cryo-bolometers. Atmosphere gives the main limitation in
sensitivity in MM wave ground based SSF experiment with
cryo-bolometers as it  will contribute up to 80{\%} in a system
temperature and in some cases may saturate (overheat) a
cryo-bolometer. The bolometer must be able to absorb up to 1 pW in
order to avoid saturation with photon NEP $\ge 2\times10^{- 17}$
Wt/Hz$^{1/2}$. This requires a rather high dynamic range D which
is the ratio of the maximum power detected to the minimal
detectable power. For expected sensitivity  1 mK s$^{1/2}$ D must
be not worse than 45 dB.  In fact only days with lowest
atmospheric opacity (lowest water vapour content) may be used for
such an experiment at sea level. It was shown by a ``tipping
curve'' method that typical atmospheric opacity at MSRT site in
zenith $\tau_{z}$ is 0.2 dB at 36.5 GHz for dry clear Autumn
nights (Khaikin et al, 2003) that gives predicted $\tau _{z}$ at
94 GHz 0.3 dB - 0.5 dB in clear cold winter nights with lowest
integrated water wapour content ($<2 g/m^{3}$). An effective beam
wobbling made it possible to reduce residual atmospheric noise
$\Delta $T$_{atm}$ and spillover $\Delta $T$_{spill}$ up to 10
$\mu$K at 40 GHz in SK CMBR experiment in Canada (Wollack et al,
1994), where the radiometric offset was less than 7 $\mu$K per day
for all observations.  A special ground screen (height up to 3 m)
is to be installed near MSRT (Fig.5) to avoid the contribution of
the hot atmosphere and ground via sidelobes, spillover and
scattering background and provide T$_{AO }$= 33 K at high
elevation angles including own antenna temperature, atmosphere (20
K) and CMBR(Khaikin, Luukanen, Dubrovich,2005). The residual
non-thermal instrumental noise must be significantly reduced up to
30 $\mu$K in differential maps while thermal noise grows by a
factor 2$^{1 / 2 }$in comparison with single frequency maps.
Non-thermal noise can not be reduced lower in difference maps  due
to residual abnormal (1/f) instrumental noise, some difference in
radiation patterns, in amplitude-frequency characteristics of the
channels and some difference in atmospheric effects within 12{\%}
bandwidth.

In spite of atmospheric limitations a ground based SSF MM wave
experiment has advantages in the cost, flexibility and possible
duration. In space conditions TES bolometers may give one - two
orders better sensitivity. In (Ali et al, 2003) NETD=25 $\mu$
K/s$^{1 / 2 }$ per pixel is expected in space for CMBR at 90 GHz
for 100 TES detectors with the physical temperature less than 0.1
K and NEP=$2\times10^{- 18}$ Wt/Hz$^{1/2}$. Yet, most of new
generation MM wave CMBR projects chose multi-element bolometer
arrays to reach highest sensitivity in the ground based
experiment. $NET_{CMBR}=260-280 \mu K s^{1/2}$ per detector  must
be reached in 1000 element array of TES bolometers at 10 m South
Pole Telescope(SPT) in 20\% bandwidth at 95 GHz and 150 GHz(Ruhl
et al, 2004). Millimeter Bolometric Array Camera(MBAC) with 3
1000-element TES bolometer arrays at 150 GHz/220 GHz/270 GHz is
under development now for 6 m  Atacama Cosmology Telescope(ACT)
Project (Fowler, 2004). Expected MBAC sensitivity to CMBR is 300
$\mu$K/500$\mu$K/700$\mu$K$ s^{1/2}$ per detector

\begin{table}
\caption{Expected characteristics of the experiment }\label{tab1}

\begin{tabular}
{|p{60pt}|p{60pt}|p{60pt}|p{50pt}|p{60pt}|}
\hline
\textsf{Alpha,} \par \textsf{hours}&
\textsf{Delta,} \par \textsf{degrees}&
\textsf{Multipol} \par \textsf{moment l}&
\textsf{Frequency } \par \textsf{band,} \par \textsf{GHz}&
\textsf{Frequency} \par \textsf{resolution, } \par \textsf{{\%}} \\
\hline
\textsf{8-16}&
\textsf{86}&
\textsf{1200-1500}&
\textsf{88-100}&
\textsf{2} \\
\hline
\textsf{Sensitivity per beam, mKs}$^{1 / 2}$&
\textsf{Integration time,} \par \textsf{s}&
\textsf{Integral sensitivity of the experiment, micro K}&
\textsf{dT/T}&
\textsf{Duration of the experiment} \\
\hline \textsf{1-2}& \textsf{170}& \textsf{50-55}&
\textsf{2$\times$10}$^{ - 5}$\textsf{}&
\textsf{Winter season} \\
\hline
\end{tabular}
\end{table}

\underline{Astrophysical Foregrounds}

In a ground based experiment at sea level atmospheric effects
contribute much more than astrophysical foregrounds nevertheless
we should take them into account to better prepare the experiment.
Galactic and extragalactic dust emission will significantly
dominate at 3 mm over other foreground effects such as discrete
and diffuse sources, galactic synchrotron and free-free emission,
hot clusters of galaxies. Dust fluctuations are seen on all
angular scales from 2$'$ to tens of degrees on IRAS data (Low et
al.,1984). Antenna temperature is $T_a \sim \nu ^{\beta _d}$,
where $\beta_d$ is the spectral index which depends upon   the
emissivity of the dust and lies in the range of 1.5-2.2  according
to COBE, IRAS and WMAP data. After 60 GHz  Galactic and
extragalactic dust dominate and we should carefully chose CMBR map
coordinates in dust free areas of the Universe at high Galactic
latitudes. The full sky thermal dust map based on data from IRAS
and COBE using MEM procedure shows a wide range of temperatures
from 400 $\mu$K to 15 $\mu$K in W band of WMAP (94 GHz) (Bennet at
el, 2004). Contribution of dust emission is to be significantly
eliminated in differential maps as this component must be well
enough correlated in single frequency maps due to its black-body
spectrum.

\section{Final remarks}

As it was shown in section 2 the expected  signal ($\Delta T/T$ of
SSF) depends on many Cosmological parameters and conditions in
Early Universe such as $\Omega_m$,$\Omega_{\Lambda}$,
$\Omega_{b}$, $n_{0H}$,$\alpha$,$V_p$ and others while the best
S/N ratio at given z also depends on $\Delta \nu/\nu$. It is
extremely difficult to predict the most probable combination of
these parameters at different z in Early Universe and SSF
experiments are urgently needed to set SSF an uppper limit at
$\Delta T/T$=$10^{-4}-10^{-5}$ level and specify theoretical
predictions of SSF.

We want to emphasize in final remarks   the significance of SSF
detection which leads to the discovery of the Cosmological
molecules. Cosmological molecules can provide an unique  test to
non-standard nuclear synthesis at high z. SSF  can only help us in
investigating Dark Ages epoch in the Early Universe ($300 > z >
10$ ). SSF detection at z=20-30 may give an important information
about ionization history of the Universe. Finally, they will allow
us to explore physical properties of Dark Matter and Dark Energy.

\section{Conclusion}

Observational effects caused by  interaction of primordial
molecules with CMBR seem to be most promising in investigating of
Dark Ages epoch of the Early Universe and ionization history of
the Universe. Strategy and concept of the experiment are given.
Simulation of the experiment in an atmospheric transparency window
at 3 mm with MSRT equipped with a 7x4 beam cryo-microbolometer
array gives us the expected sensitivity of 50-55 $\mu$K in the
differential CMBR maps near NCP with angular resolution 6$'$-7$'$
(l=1200-1500) and frequency resolution \textbf{$\Delta $}$\nu
$/$\nu $=2{\%}. The expected integral sensitivity in the
experiment is $\Delta $T/T=2$\cdot $10 $^{\mbox{--}5}$ which is
near to optimistic estimates of SSF caused by molecules HeH$^{+}$
at z=20-30. Simulation shows that SSF may be recognized in the
angular power spectrum when S/N in single frequency CMBR maps is
as small as 1.17  for a white noise. Such an experiment will give
us a possibility to set upper limits in SSF observations in MM
band and prepare future SSF experiments.

\label{lastpage}

\begin{thebibliography}{}

\bibitem{1}

Ali,~L.~et~al.~Planar~Antenna-Coupled T ransition-Edge Hot
Electron Microbolometer, 2003, IEEE TRANS. AS, 13.

\bibitem{2}

Basu, K., Hernandez-Monteagudo, C., Sunyaev, R., 2004, A\&A, 416,
447-466.

\bibitem{3}
Benford, D. et al., A planar two-dimensional  superconducting
bolometer array for the Green bank Telescope, 2004, SPIE, 5498,
208-218.

\bibitem{4}


Bennet, C.L, Hill R.S., G.Hinshow et al., 2004,  ApJ, in press.

\bibitem{5}
Crill B.P. et al., 2002, Ap J, v.676.

\bibitem{6}
de Bernardis, P., V.Dubrovich et al., 1993, A\&A, 269, issue 1/2,
1-6.

\bibitem{7}

de Bernardis et al., 2002, Ap.J., 564, 559.

\bibitem{8}
Bond, J.R., Efstathiou, G., 1987, MNRAS, 226, 655.

\bibitem{9}

 Doroshkevich, A. G., Dubrovich, V. K., 2001, MNRAS, 328, 79.


\bibitem{10}

Dubrovich, V.K., 1975, Soviet Astron. Letters, 1, 196.

\bibitem{11}


Dubrovich, V.K., 1977, Astron. Letters, 3, 243.

\bibitem{12}

Dubrovich, V.K., 1977/2, PhD thesis, St.Petersburg, Pulkovo.

\bibitem{13}

Dubrovich, V.K, 1982, Izvestia SAO, 15, 21.

\bibitem{14}

Dubrovich, V.K., Lipovka, A.A., 1995, A\&A, 296,301.

\bibitem {15}

Dubrovich,V.K., Stolyarov, V.A., 1995, A\&A, 302, 635.


\bibitem{16}


Dubrovich, V.K., 1997, A\&A, 324, 27.

\bibitem{17}

Dubrovich, V.K., Bajkova, A.T., 2003, Astron. Lett., 29, 567-572.

\bibitem{18}

Dubrovich, V.K.,Bajkova, A.T., Khaikin, V.B. In Proceedings  of
the International Conference "Exploring the Cosmic Frontier:
Astrophysical Instruments for the 21 st. Century", Berlin,
Germany, 2004.

\bibitem{19}
Dubrovich, V.K., Shakhvorostova, N.N., 2004, Astron. Lett.,
30,509.

\bibitem{20}

Dubrovich, V.K., Grachev, S.I. 2004, Astron. Lett., 30, 657.

\bibitem{21}

Goldin, A., Bock, J. et al., Design of broadband filters and
antennas for SAMBA, 2003, in Proceed. Of SPIE, 4855, 163-171.

\bibitem{22}

Fixsen, D.J., Mother, J.C., 2002, ApJ, 581, 817.

\bibitem{23}

Fixsen et al., 2004, ApJ, 612, 86-95.

\bibitem{24}
Gorski,K.M., 1997, In Proceedings of the XXXI-st Recontres  de
Marion Astrophysics Meeting, 77.

\bibitem{25}
Granet, C.,James, G.L., Bolton, R., Moorey, G. 2004,  Antennas and
Propagation, IEEE Trans., 52, 848 - 854.

\bibitem{26}

Fowler,J. The Atacama Cosmology Telescope Project, 2004, SPIE,
5498.

\bibitem{27}

Gosachinskii, I.V., Dubrovich, V.K. et al., 2002, Astron. Rep.,
46, 543-550.

\bibitem{28}

Kasyanov, A., Khaikin, V., In Proceedings of 27  th ESA Antenna
Technology Workshop on Innovative Periodic Antennas (WPP-22),
Santiago de Copostela, Spain, March 9-11, 2004.

\bibitem{29}

Khaikin, V., Yakovlev, S., Kazarinov A., Efimov,  I., Volkov A.,
Valtaoja E. Convention on Radio Science, pp.84-87, Espoo, Finland,
Oct.2002.

\bibitem{30}

Khaikin,V., Yakovlev S., Kazarinov A., Karavaev,D.,  Rybakov, Yu.
2003, In Proceed. of 3-d ESA Workshop on MM-wave Technology and
Applications, Espo, Finland, 425-430.

\bibitem{31}

Khaikin, V., Luukanen A. 2003, In Proceed.of   3-d ESA Workshop on
MM-wave Technology and Applications, pp.419-425, Espo, Finland.

\bibitem{32}
Khaikin, V.B., Dubrovich, V.K. MM-wave radio  telescope with
multibeam focal plane array for SSF search. 2004, in Proceedings
of IRMMW-2004, Karlsruhe, Germany.

\bibitem{33}

Khaikin, V.B., Luukanen, A., Dubrovich, V.K.,  Instrumental and
atmospheric noise in ground based MM-wave SSF experiment, 2005,
G\&C, 11.


\bibitem{34}
Khaikin, V.B. Simulation of antenna characteristics  of mm-wave
radio telescopes with multibeam FPA. In Proceedings of 28 th ESA
Antenna Workshop on Space Antenna Systems and Technologies,
Nordwijk, The Netherlands, May, 2005.

\bibitem{35}

Khaikin, V.B., Chung M-H., Radzihovsky V.N., Kuzmin S.E.,  Kaplya
S.V., Simulated characteristics of 14 m MM-wave TRAO telescope
with multibeam FPA at 85-115 GHz, 2005, G\&C, 11.


\bibitem{36}
Khaikin, V.B., Radzikhovsky, V.N. , Kuzmin S.E., Shlenzin, S.V.,
Popenko, Chernobrovkin, R. Granet, C. Tightly packed multibeam FPA
at 30-38 GHz. In Proceed. of 29 th ESA Antenna Conference,
Noordwijk, The Netherlands, April, 2007.


\bibitem{37}

Kogut A., Spergel, D.N., Barnes, C. et al., 2003, ApJS, 148, 161.

\bibitem{38}

Lawrence, C.R., Gaier, T., Seiffert, M., Millimeter-Wave  MMIC
Cameras and the QUIET Experiment, 2004,. Proceedings of SPIE,
5498.

\bibitem{39}
Lepp, S., Shull, J.M., 1984, ApJ, 280, 465.

\bibitem{40}


Low, F.J., Beintema, D.A., Gautier, T.N., Gillett,  et al. 1984.
Ap. J. Lett. 278; L19-L22.

\bibitem{41}


A.Luukanen, J.Pekola. Applied Physics Letters, v.82,
pp.3970-3972, 2003.


\bibitem{42}
Maoli, R., Melchiorri. F., Tosti, P. 1994, ApJ., 425, 372-381.

\bibitem{43}

Maoli, R. et al., 1996, ApJ., 457, 1.

\bibitem{44}
Majorova, E.K., Khaikin, V.B. Characteristics of radio  telescopes
with multielement microstrip focal plane arrays. Radiophysics,
v.XLVIII, N2, pp.95-109, 2005.

\bibitem{45}

Mather, J.C., Cheng, E.S. et al., ApJ, 420,439. 1994.

\bibitem{46}

Mauskopf, P.D. B.K. Rownd, S.F. Edgington, V.V. Hristov,  A. K.
Mainzer, J.Glenn, A.E. Lange, J.J. Bock, and P. A. R Ade. Proc.
Imaging at Radio through Submillimeter Wavelengths. 2000, ASP
Conference Seriess, 217, p.115.

\bibitem{47}

Miller, A.D. Proceedings of MG-9, World Scientific,  Singapore,
2001.


\bibitem{48}
Myers M. et al. Applied Physics Letters, 86,114103, 2005.

\bibitem{49}


Page, L. Millimeter-submillimeter wavelength filter  system.1994,
Applied Optics.v.33, N1.

\bibitem{50}
Page, L. et al. Astro-ph/0603450, 2006.

\bibitem{51}
Palla,F et al.,1995, ApJ,451,44.

\bibitem{52}

Parijskij, Y.N., Korolkov, D.V. Experiment "Cold" ,  Sov. Sci.
Rev. E Astrophys. Space Phys., Vol. 5, 1986, pp. 39-179.

\bibitem{53}

Parijskij, Yu.N. Bursov, N.N. Berlin, A.B.  Balanovskij,  A.A.
Khaikin, V.B.  Majorova, E.K.  Mingaliev, M.G.  Nizhelskij, N.A.
Pylypenko, O.M. Tsibulev, P.A. Verkhodanov, O.V. Zhekanis, G.V.
Zverev,Yu.K. RATAN-600 new zenith field survey and CMB problems.
Gravitation\&Cosmology, 10(2004), pp.1-10.

\bibitem{54}

Peacock,J.A.1999, Cosmological Physics,  Cambridge Univ.Press,
chap.18.

\bibitem{55}

Peebles, P.J.E.1993, Principles of Physical Cosmology,  Princeton
Univ.Press, chap.21.

\bibitem{56}


Puy, D. et al., 1993, A{\&}A, 267, 337.


\bibitem{57}

Readhead, A. et al. 2004, ApJ, 609, 418.

\bibitem{58}

Rubino-Martin,J.A, Mernandez-Monteagudo, C. and  Sunyaev, R.A.,
2005, Astro-ph/0502571

\bibitem{59}

Ruhl, J.E. et al. The South Pole Telescope. 2004,  Proc. SPIE,
5498.

\bibitem{60}

Spergel, D.N., Verde, L., Peris, H.V. et al., 2003, ApJS, 148,
175.

\bibitem{61}

Spergel, D.N. et al., 2006, Astro-ph/0603449.

\bibitem{62}

Stancil, P.D., Lepp, S., Dalgarno, A., 1996, ApJ, 458, 401S.

\bibitem{63}

Sunyaev, R.A., Zel'dovich Ya.B., 1970. Ap{\&}SS, 7, 3-19.

\bibitem{64}

Sahni, V.,Starobinsky, A., 2000, Int.J.Mod.Phys., 9, 373-444.

\bibitem{65}

Varshalovich, D., Khersonskii, V.,  Suyaev, R., 1981,
Astrophysics, 17, 273.

\bibitem{66}

Vdovin, V.F. Radiophysics, 2005, 48, 876.

\bibitem{67}

Wyithe,J., Loeb, A., 2003, ApJ, 322, 597.

\bibitem{68}

Wollack, E. J., Jarosik, N., Netterfield et al.,1 994,  Astrophys.
Lett. Commun., 35, 217.

\bibitem{69}
Zel'dovich, Ya.B., Novikov, I.D.,  Structure and evolution of the
Universe, Moscow, Nauka, 1975.


\bibitem{70}

Zel'dovich, Ya.B., 1978, Sov.Astron.Lett., 4, 165.

\end{thebibliography}
\end{document}